# A Tipping Pulse Scheme for a rf-SQUID Qubit

Xingxiang Zhou, Jonathan L. Habif, Andrea M. Herr, Marc J. Feldman, Mark F. Bocko

*Abstract*— We present a technique to control the quantum state of a rf-SQUID qubit. We propose to employ a stream of single flux quantum (SFQ) pulses magnetically coupled to the qubit junction to momentarily suppress its critical current. This effectively lowers the barrier in the double-well rf SQUID potential thereby increasing the tunneling oscillation frequency between the wells. By carefully choosing the time interval between SFQ pulses one may accelerate the interwell tunneling rate. Thus it is possible to place the qubit into a chosen superposition of flux states and then effectively to freeze the qubit state. We present both numerical simulations and analytical time-dependent perturbation theory calculations that demonstrate the technique. Using this strategy one may control the quantum state of the rf SQUID in a way analogous to the π pulses in other qubit schemes.

## INTRODUCTION

It has been recognized for at least two decades that the rf-SQUID may present an opportunity to observe quantum coherent behavior in a macroscopic system [1]. Specifically, evidence for the superposition of macroscopically distinct quantum states in a SQUID, à la Schroedinger's cat, has long been sought [2,3]. Such quantum superposition is the essence of quantum bit (qubit) behavior in a quantum computer and about four years ago the rf-SQUID was proposed as a candidate to serve as the qubit for a possible superconducting quantum computer [4]. Subsequently another Josephson junction circuit, the so called persistent current qubit, was also proposed [5]. Very recently evidence for the existence of the superposition of macroscopically distinct states in the rf SQUID [6] and in the persistent current qubit [7] has been presented. Although the macroscopic quantum coherent oscillations characteristic of the quantum superposition of energy eigenstates of the rf SQUID have yet to be observed the recent experiments [6,7] have displayed the avoided crossing of two energy levels in the system which is evidence for quantum superposition and encouraging news for possible superconducting quantum computer realizations.

In this paper we consider the next step in the development of a superconducting quantum computer - how to control the quantum state of the qubit. We focus on the rf SQUID qubit and present a technique employing RSFQ control circuits to manipulate superpositions of the two quantum states used as the rf SQUID qubit basis states. This provides a level of control for the SQUID qubit akin to the rotation of atomic spin qubits [8]. Moreover RSFQ electronic circuits allow picosecond timing accuracy and the convenience of on-chip control – a feature that we believe to be critical for the eventual development of a practical quantum computer.

For properly chosen parameters the rf SQUID provides a two-state system that may exhibit the quantum behavior necessary to realize a qubit. In our proposed scheme, the rf-SQUID is biased at one-half flux quantum so that the potential is a symmetric double well. The SQUID loop inductance and junction critical current are chosen such that there are four energy levels below the potential barrier height. The first two energy levels, of even and odd parity, and separated by about 50 MHz, provide the qubit states. The superposition of the lowest two energy levels yields a "flux state" for which the wavefunction oscillates between the two potential wells at the beat frequency of the two energy levels. At certain times the wavefunction is localized in one or the other of the potential wells with the expectation value of the total flux linking the SQUID being somewhat greater than or less than one-half flux quantum. This corresponds to a macroscopic current in the SQUID loop circulating in one or the other direction.

If the state of the system is localized in one of the two potential wells, corresponding to the equal amplitude superposition of the first two energy eigenstates, and then allowed to evolve freely, the probability of finding the system in either one of the wells is a sinusoidal function of time. Initial state preparation may be accomplished by making a measurement of the SQUID flux. The required measurement interaction is straightforward to realize and the corresponding classical measurement is routinely made. Furthermore, using RSFQ circuits to perform the flux measurement allows us to make a very fast, i.e., adiabatic, measurement of the flux. Thus the significant problem of the initial state preparation is apparently solved. Since the first two energy levels are nearly degenerate for the chosen rf SQUID parameters, the coherent oscillation of the flux between the two wells proceeds on a relatively slow time scale, on the order of tens of nanoseconds or longer. Under this condition the rf SQUID may be considered to be in a quasi-static "flux" state for the time-scales of interest. In our proposed scheme we employ the magnetic field generated by an SFQ pulse propagating on a transmission line magnetically coupled to the rf SQUID Josephson junction to suppress its critical current and thereby lower the central barrier of the two well potential. This effectively accelerates the sinusoidal oscillation of the flux between the two wells. By applying a sequence of SFQ pulses with carefully timed inter-pulse spacing we may control the oscillation of the flux in the rf SQUID and bring its state to any linear combination of the two flux states. At the end of the SFQ pulse train the potential barrier reverts to its original height and the system is effectively frozen in the superposition that resulted from the interaction with the SFQ pulse sequence. This is our proposed means of rotating the

Manuscript received September 19, 2000. This work was sponsored by the Army Research Office, NSA and ARDA under Grant No. DAAG55-98-1-0367.

Authors are with the Department of Electrical and Computer Engineering, University of Rochester, Rochester, New York 14627. (email bocko@ece.rochester.edu)

A.M. Herr was with University of Rochester. She is now with Sierra Monolithics, Inc., 103 West Torrance Blvd., Redondo Beach, CA 90277.

state vector of the SQUID qubit. Below we present numerical simulations and an analytical analysis for our "tipping pulse scheme".

## HAMILTONIAN OF THE RF-SQUID

The potential energy of the rf SQUID as a function of the total flux linking the SQUID loop is well known. In the case of interest here we choose the external flux bias, $\phi_x$ to be exactly one-half flux quantum, $\phi_0/2$, creating a symmetric potential well with two equal energy minima. The Hamiltonian for the rf SQUID is:

$$H_0 = -\frac{\hbar^2}{2C}\frac{d^2}{d\phi^2} + \frac{(\phi-\phi_x)^2}{2L} - \frac{I_c \phi_0}{2\pi}\cos(2\pi\frac{\phi}{\phi_0}) \quad (1)$$

where C is the capacitance of the Josephson junction, L is the SQUID inductance, and $I_c$ is the SQUID junction critical current. One may view the rf SQUID as a quantum mechanical particle of mass C moving in the double well potential. The practical realization of our rf SQUID qubit will employ a very small double junction SQUID in place of the single junction of the idealized rf SQUID. This well known technique allows one to adjust the critical current of the "junction" by controlling the amount of magnetic flux linking the double junction SQUID loop. In practice the critical current $I_c$ as a function of the flux linking the double junction SQUID is $I_c = I_{c0}|\cos(\phi_J/\phi_0)|$ where $I_{c0}$ is the critical current of the double junction SQUID in the absence of any applied flux and $\phi_J/\phi_0$ is the flux linking the double junction SQUID normalized to the flux quantum, $\phi_0$. In practice, if the inductance of the double junction SQUID is a small fraction of the rf SQUID inductance then the dynamics of the "adjustable critical current" rf SQUID are nearly identical to those of the ideal rf SQUID constructed with a single junction. We plan to control the critical current by using a microstrip line located above the double junction SQUID down which SFQ pulses may be propagated. In practice we need only suppress the critical current by 1% to 2%.

Intuitively, lowering the potential barrier increases the tunneling rate between the two wells, and the oscillation of the wavefunction between the two potential minima speeds up. However, the details of the dynamics are more complicated, and the upper energy levels are found to play an important role. The simple W. K. B approximation does not apply in this case and there are two different ways to address this problem. In the first, one may solve Shroedinger's equation for the two different potentials, corresponding to the two values of $I_c$ to obtain two sets of (slightly) different eigenenergies and eigenfunctions. Since the wave function evolves with different phase advance rates during and in-between the SFQ pulses, one can project the wavefunction back and forth onto the two sets of basis states. This works if we make the assumption that the critical current is a block wave in time (so there is a well defined set of eigenfunctions during the intervals during which the critical current is suppressed). Alternatively, one can treat the change in critical current as a perturbation and apply time-dependent perturbation theory.

## NUMERICAL SIMULATION

For simplicity we idealize the stream of SFQ pulses serving to suppress the SQUID junction critical current as a block function. The duration of a SFQ pulse is $t_d$ and the space between SFQ pulses is $t_s$. The numerical simulation consists of the following: first, one must calculate the eigenfunctions and energy eigenvalues in the two potentials that correspond to the two slightly different critical currents; then one chooses an initial wavefunction, usually the equal superposition of the lowest two energy levels in the unperturbed potential. The next step is to propagate the wavefunction forward in time in the unperturbed potential until the time when the first SFQ pulse lowers the SQUID critical current. At this instant the current wavefunction is projected onto the eigenfunctions of the new potential; normally we need to only consider the first 10 eigenfunctions. After finding the representation of the wavefunction in the perturbed basis states we may once again propagate the wavefunction forward in time for a time $t_d$, while the SFQ pulse interacts with the SQUID junction. At the end of the interval the wavefunction is projected back onto the unperturbed basis states and the entire process is repeated as indicated by the SFQ pulse stream.

In Fig. 1 we show the result of the simulation for one particular choice of the SFQ pulse stream timing. In the case shown we assume that $t_d$ is 3 psec and that $t_s$ is 25.9 psec. It is assumed that the SQUID junction critical current is suppressed by 1% - the other assumed parameters are given in the figure caption. The first four energy levels are the most important in the dynamics of the system when the perturbation is small. The result of the perturbation is to populate the 3rd or 4th energy levels transiently, eventually transferring all of the probability back into the first two energy levels. Due to the symmetry of the wavefunctions the $1^{st}$ and $3^{rd}$ energy levels (which are both symmetric) are coupled to one another and the antisymmetric, $2^{nd}$ and $4^{th}$, levels are also coupled. In the figure the probability of occupation of each of the first four energy levels is plotted versus time. Also plotted is the probability of finding the system in the right or left hand potential wells versus time. For the specific choice of inter-pulse timing there is large transfer of probability between the 1st and 3rd energy levels. For a slightly different value of inter-pulse timing, $t_s$ = 23.9 psec, a strong interaction between the 2nd and 4th energy levels arises and a similarly appearing plot results.

The key to speeding up the coherent oscillation of the flux is to excite the system to one of the higher energy levels, in which the phase of the wavefunction evolves more rapidly. Then after a specific time interval has elapsed one must transfer all the probability back to the lowest two energy levels. Since the tunneling rate (and the resulting inter-well oscillation) is very slow for the unperturbed system this enables one to bring the flux states, our qubit basis states, into any chosen superposition and then to "freeze" the state of the system by reverting to the unperturbed potential.

In the simulation result presented above the natural MQC oscillation period was fairly short - approximately 10.5 nsec. The sequence of critical current suppressing RSFQ pulses sped this up to approximately 1.7 nsec. This is the length of time that it takes to affect a π pulse for the rf SQUID qubit in this illustration. Of course, one would like the natural MQC oscillation frequency to be as low as possible and the π-pulse duration to be as short as possible. We are presently searching for parameters that achieve this goal.

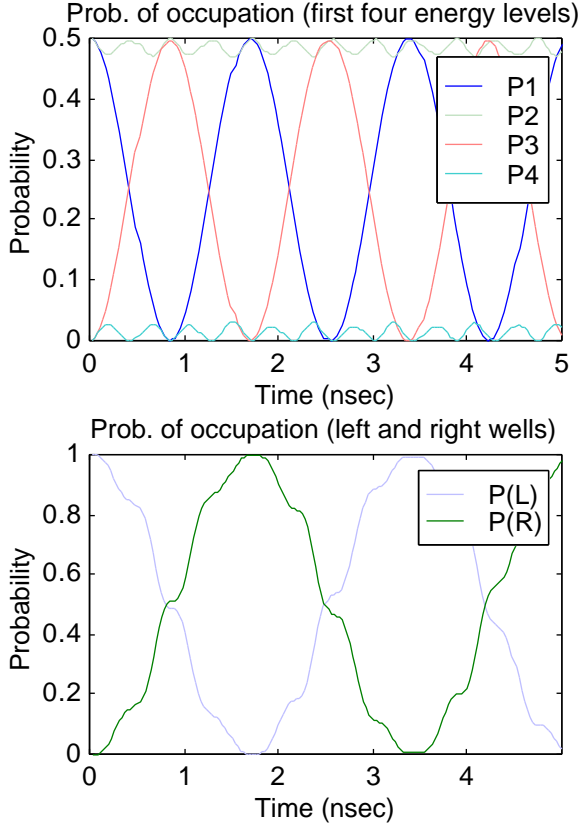

Figure 1 Results of the numerical simulation of the rf SQUID interacting with a SFQ pulse stream. The SQUID parameters are: L = 97 pH, unperturbed critical current $I_c$ = 4 μA, and C = 50 fF. The assumed reduction of the critical current is 1%, to $I_c'$ =3.96 μA. The SFQ pulse stream timing parameters are $t_d$ = 3 psec (duration of a pulse) and $t_s$ = 25.9 psec (interpulse spacing).

TIME-DEPENDENT PERTURBATION THEORY CALCULATION

To gain more insight into the underlying physics, we solved Schroedinger's equation analytically using time-dependent perturbation theory. We maintain the assumption that the critical current as a function of time is a block wave. In addition, we focus our attention on the first four levels. This approximation has the virtue of allowing us to solve the equations exactly but to still grasp the most important aspects of the dynamics. For the "on-resonance condition" discussed below, we will see that for certain choices of the timing parameters, this truncation of the entire (infinite) Hilbert space is justified.

The profile of the critical current suppression is characterized by three parameters: the amount of the critical current suppression ($I_c'$-$I_c$), the duration of the suppression $t_d$, and the interval between successive suppressions, $t_s$. We write the Hamiltonian as (in the shifted flux coordinate $\phi - \phi_0/2$) as $H = H_0 + H'$ where $H_0$ is the unperturbed Hamiltonian as shown above (corresponding to the unsuppressed critical current $I_c$), and $H'$ is the perturbation Hamiltonian,

$$H' = \frac{I_c' - I_c}{2\pi} \phi_0 \cos(2\pi \frac{\phi}{\phi_0}) \quad . \tag{2}$$

Note that H' is an even function of the flux. Because of this symmetry, the matrix elements of H' between states with opposite parity vanishes: <i|H'|j> = 0 if the wavefunctions of states i and j have different symmetries. In this case states i and j are decoupled, and transitions between them are forbidden. This is the selection rule for allowed SQUID transitions. Since levels 1 and 3 are even, and levels 2 and 4 are odd, the H matrix is:

$$H = \begin{pmatrix} E_1 + H_{11} & 0 & H_{13} & 0 \\ 0 & E_2 + H_{22} & 0 & H_{24} \\ H_{31} & 0 & E_3 + H_{33} & 0 \\ 0 & H_{42} & 0 & E_4 + H_{44} \end{pmatrix} \tag{3}$$

If we write the wavefunction as

$$\psi(t) = \sum_{i=1}^{4} a_i(t) e^{iE_i t/\hbar} |i\rangle \tag{4}$$

and substitute it into Schroedinger's equation, we get two sets of decoupled equations. The first set is

$$i\hbar \frac{d}{dt} a_1 = H_{11} a_1 + H_{13} e^{i(E_1 - E_3)t/\hbar} a_3 \\ i\hbar \frac{d}{dt} a_3 = H_{31} e^{i(E_3 - E_1)t/\hbar} a_1 + H_{33} a_3 \tag{5}$$

the second set of equations can be found by making the substitution 1—2, 3—4 in Eqns. 5. Since the two sets of equations are the same, we need only solve one set, say the equations for $a_1$ and $a_3$. We let

$$a_1 = b_1 e^{-iH_{11} t/\hbar} \quad ; \quad a_3 = b_3 e^{-iH_3 t/\hbar} \quad , \tag{6}$$

which allows us to find the two $2^{nd}$ order differential equations obeyed by $b_1$ and $b_3$. The equation for $b_3$ is

$$\frac{d^2}{dt^2} b_3 - i\frac{\lambda}{\hbar} \frac{d}{dt} b_3 + \left|\frac{H_{13}}{\hbar}\right|^2 b_3 = 0 \quad . \tag{7}$$

If we assume that the $3^{rd}$ level is initally unpopulated i.e., $b_3(0) = 0$, the solution is:

$$b_3(t) = A e^{i\lambda t} \sin(\nu t) \tag{8}$$

where

$$\lambda = (H_{33} - H_{11} + E_3 - E_1)/2\hbar$$

$$\nu = \sqrt{(H_{33} - H_{11} + E_3 - E_1)^2 + 4|H_{13}|^2}/2\hbar \tag{9}$$

$$A = \frac{1}{\nu} \frac{H_{31}}{i\hbar} b_1(0)$$

Thus the population in level 3 responds as a function of the perturbation. For $b_3$ to eventually return to zero, we require $t_s$ = n$\pi$. Unfortunately, even the minimum achievable time for $t_s$ is much longer than the duration of the perturbation (that of a single SFQ pulse). Therefore a sequence of SFQ pulses with carefully chosen timing must be used. It is clear that each new pulse must arrive when the phase relation between $a_1$ and $a_3$ returns to the value where it "left off" after the previous pulse. In this way the phase $\nu t$ increases continuously and the amplitude of $b_3$ follows a sinusoidal dependence. This on-resonance condition for the inter-pulse spacing $t_s$ is $t_s = mh/(E_3-E_1)$ where m is any integer. Obviously the same analysis can be completed for $a_2$ and $a_4$, with $H_{13}$ replaced by $H_{24}$. From these results we may compute the an approximate formula for the phase advance between $a_1$ and $a_2$ throughout one pulse stream cycle,

$$\theta \approx \frac{1}{2}(\omega_4 - \omega_3)t_2 + \frac{1}{2}(\omega_2 - \omega_1)t_2 \quad . \quad (10)$$

Instead of the macroscopic quantum coherent oscillation proceeding at its unperturbed characteristic frequency, $(\omega_2-\omega_1)/2$, during the SFQ pulse perturbation the oscillations proceed at the frequency $(\omega_4-\omega_3)/2$ which may be orders of magnitude faster than $(\omega_2-\omega_1)/2$. It is in this way that the upper energy levels 3 and 4 play an essential role in the dynamics.

It is obvious that the above "resonance" conditions can not be satisfied simultaneously for $(a_1, a_3)$ and $(a_2, a_4)$. We are limited to a choice of $t_d$ and $t_s$ such that one pair of states is in resonance, say $a_1$ and $a_3$. In that case there is little excitation from $a_2$ to $a_4$. The numerical results reported above also demonstrate this behavior.

## SUMMARY

If a practical quantum computer can be built it will offer an entirely new form of information processing far beyond the capabilities of classical computation. The unique attributes of superconductors make it likely that practical solid-state quantum computers will employ superconducting circuits in some form, either to serve as the qubits, control electronics or both.

In this paper we addressed the problem of how to achieve quantum control of a superconducting qubit. That is, given a set of qubit basis states, the flux states of the rf SQUID in our case, how may one drive the state of the system into any chosen superposition of the basis states? We have shown that, in principle at least, one may accomplish this by employing controlled interactions with single flux quantum pulses. The simulations and simplified analytical model presented here allow us to better understand the nature of the quantum control process in the rf SQUID. We see that the interaction is a type of parametric process, wherein a parameter of the qubit, the junction critical current, is modulated in a time dependent fashion. In this process energy is "pumped" into the qubit, (the upper energy levels are excited) and then extracted again at a specific time with the net effect of leaving the higher energy levels unpopulated. A related recent theoretical study presents a method employing dynamic pulse control techniques to cancel transitions to higher energy levels arising in the interactions among multi-level qubits [9]. In our scheme we take advantage of the higher energy levels by effectively speeding up the dynamic evolution of the qubit, i.e., the oscillation of the flux between the two potential wells. Upon deexcitation of the qubit, back to the first two energy levels, the superposition state of the qubit is effectively "frozen". By carefully orchestrating this procedure one may construct an arbitrary superposition of qubit basis states.

We have presented a preliminary step to achieving quantum control of superconducting qubits. We believe that the basic principles presented here will lead to much more refined qubit control procedures and that the use of RSFQ control circuits also will likely find applications in a much wider class of qubit circuits.